\documentclass[cits,proceedings]{PoS}
\usepackage{amssymb,amsmath}
\usepackage{mathbbol}

\newcommand{\Umu}{U_{\mu}}

\title{Hunting the static energy renormalon}

\ShortTitle{Hunting the static energy renormalon}

\author{\speaker{Clemens~Bauer}, Gunnar Bali\\
        Institut f\"ur Theoretische Physik, Universit\"at Regensburg\\\
93040 Regensburg, Germany\\
        E-mail: \email{clemens.bauer@physik.uni-regensburg.de}\\
        \hspace*{1.2cm}\email{gunnar.bali@physik.uni-regensburg.de}}

\abstract{We employ Numerical Stochastic Perturbation Theory (NSPT) together with twisted boundary conditions (TBC) to search for the leading renormalon in the perturbative expansion of the static energy. This renormalon is expected to emerge four times faster than the one for the gluon condensate in the plaquette. We extract the static energy from Polyakov loop calculations up to $12~$loops and present preliminary results, indicating a significant step towards confirming the theoretical expectation.}

\FullConference{The XXVIII International Symposium on Lattice Field Theory\\
         June 14-19,2010\\
         Villasimius, Sardinia, Italy}

\begin{document}
\section{Motivation}
It is known since long that QCD perturbation theory is divergent:
at best, the perturbative coefficients form an asymptotic series.
The coefficients $k_n$ of a generic expansion,
\begin{equation}
K=\sum_n k_n\alpha^n\,,
\end{equation}
will diverge at least like $a_d^nn!$, with a constant $a_d$
(see ref.~\cite{Beneke:98} for a comprehensive review).
This pattern of factorial growth can be inferred from
combinatorial studies of the contributing Feynman diagrams
and is related to the position of the first renormalon pole
in the complex Borel plane.
Successive contributions~$k_n\alpha^n$ decrease for small
orders $n$ down to a minimum at $n_0\sim1/(|a_d|\alpha)$.
Higher-order contributions
should be neglected and introduce an ambiguity of the order
of this minimum term,
$k_{n_0}\alpha^{n_0}\sim\exp[-1/(|a_d|\alpha)]$.
Integrating the one-loop QCD $\beta$-function from a momentum
scale $q$ down to a cut-off parameter $\Lambda\ll q$ one obtains,
\begin{equation}
\label{eq:renormal}
\left(\frac{\Lambda}{q}\right)^d=\exp\left(-\frac{1}{|a_d|\alpha}\right)\,,\quad\mbox{where}\quad |a_d|=\frac{2\beta_0}{d}\,.
\end{equation}

The above similarity of expressions is not accidental.
Within the operator product expansion (OPE),
observables $R$ can be factorized into short-distance Wilson coefficients $C_i(q,\mu)$ and non-perturbative
matrix elements $\langle O_i(\mu,\Lambda)\rangle$ of dimension $i$:
\begin{equation}
R(q,\Lambda)=C_0(q,\mu)\langle O_0(\mu,\Lambda)\rangle
+C_d(q,\mu)\langle O_d(\mu,\Lambda)\rangle\left(\frac{\Lambda}{q}\right)^d+\cdots\,.
\end{equation}
$\mu$ denotes the matching scale, $q$ is a perturbative and $\Lambda$ a low
momentum scale so that $q\gg\mu\gg\Lambda$.
For the plaquette, $\langle O_0\rangle=1$ and the next higher non-vanishing operator is the dimension
$d=4$ gluon condensate. In this case, the perturbative expansion of $C_0$ cannot be more accurate
than $(\Lambda/q)^4$ which is exactly of the size of $k_{n_0}\alpha^{n_0}$, see eq.~(\ref{eq:renormal}): the so-called
leading infrared renormalon
of this expansion cancels the ultraviolet ambiguity of the next order non-perturbative matrix element
so that the physical observable $R$ is well defined.

Here we investigate the renormalon of the perturbative
expansion of the static energy. In this case $d=1$ which
means that we expect this expansion to start diverging at
an order $n_0$ that amounts to about one fourth of
that for the plaquette. Moreover, the ratios of two subsequent
coefficients should asymptotically be larger
by this same factor since the position of 
the first singularity
in the Borel plane is four times closer to the origin
($u=d/2=1/2$ instead of $u=2$).
QCD renormalon studies are particularly interesting because more and more diagrammatic
three-loop~\cite{Anzai:2010td}
(and even four-loop~\cite{Ritbergen:97}) calculations become available so that
an extrapolation of these existing results to even higher orders may be feasible if
the Borel structure is understood.

High-order perturbative expansions in lattice regularisation were made possible by
numerical stochastic perturbation theory (NSPT)~\cite{DRMMOLatt94,DRMMO94},
and the renormalon study of the plaquette was its first application.
Below we will describe the basic elements of NSPT, introduce twisted boundary conditions
that we employ and present first results on the static energy renormalon.

\section{Numerical stochastic perturbation theory}
NSPT is based on stochastic quantization \cite{PaWu}. We first explain the concept
for a scalar field~$\phi(x)$ and an action~$S[\phi]$.
One introduces an additional, totally fictitious stochastic time~$t$.
The evolution of the field~$\phi$ in stochastic time is dictated by a Langevin equation,
 \begin{equation}
 \frac{\partial\phi(x,t)}{\partial t}=
  -\frac{\partial S[\phi]}{\partial\phi}+\eta(x,t)\,,
\end{equation}
where $\eta(x,t)$ is a Gaussian noise. In order to calculate a generic
observable~$R$,
stochastic quantization postulates the equivalence of ensemble time averages,
in the limit of infinite stochastic time,
\begin{equation}
 Z^{-1}\!\!\int [D\phi]\,
 R[\phi(x)]e^{-S[\phi(x)]}=\lim_{t\rightarrow\infty}
 \frac{1}{t} \int_0^t \! {\rm d} t' \,
 \big\langle R[\phi(x,t')]\big\rangle_{\eta}\;.
\end{equation}
In lattice QCD, the Langevin equation must be formulated such that
the gauge links~$\Umu$ evolve within the group. This can be achieved by defining~\cite{DR0},
\begin{equation}
\label{eq:LangevinLink}
 \partial_{t}U_{\mu}(n,t) = -it^A\Bigl( \nabla_{\!n,~\!\mu,~\!A} S[U]+\eta_{\mu}^A(n,t) \Bigr)
 U_{\mu}(n,t) \;,
\end{equation}
where~$t^A$ are the generators of the $\mathrm{su}(3)$ algebra,
$\nabla_{\!n,~\!\mu,~\!A}$~is a left Lie derivative and~$\eta_{\mu}^A(n,t)$ constitute the components of
the Gaussian noise. Perturbation theory comes into play 
when rewriting each link~$U$
as a series:
\begin{equation}
 U = \mathbb{1}+\beta^{-\frac{1}{2}} U^{(1)} + \beta^{-1} U^{(2)} + \dots\,,\qquad \beta^{-1}=\frac{g_0^2}{6}=\frac{2\pi\,\alpha}{3}\,.
\end{equation}
Inserting this series into the Langevin equation eq.~\eqref{eq:LangevinLink}, one obtains a hierarchical system
of differential equations where a given order only depends on the preceeding lower orders. The perturbative series can be truncated at any desired order~$m$. NSPT is the numerical implementation
of this concept, with a discretized stochastic time~$t$
within eq.~\eqref{eq:LangevinLink}.
This necessitates simulations at different time steps~$\Delta\,t$, with a subsequent extrapolation towards~$\Delta\,t=0$.
Here we employ a second-order integrator~\cite{Torrero:08}.
We point out that the computer time naively scales like~$m^2$,
which clearly favors NSPT over diagrammatic approaches in the region of large~$m$.
\section{Twisted boundary conditions}
So far in NSPT only periodic boundary conditions (PBC) have been employed. In this case
zero modes need to be subtracted, for instance after each Langevin update.
However, one can equally well impose twisted boundary conditions (TBC) \cite{tHooft,Parisi,LuscherVertex,Arroyo}.
We assume a lattice of dimension~$L^4$. The TBC are defined by
constant twist matrices~$\Omega_\nu\in\mathrm{SU}(3)$:
\begin{equation}
   \Umu(x + L \hat\nu) = \Omega_\nu \Umu(x) \Omega_\nu^\dagger\,.
\label{Utwist}
\end{equation}
The twist matrices satisfy the relations,
\begin{equation}
   \Omega_\mu \Omega_\nu = \eta \Omega_\nu \Omega_\mu\,,\quad\mbox{where}\quad
   \eta= \exp\left(\frac{2\pi ik}{3}\right)\,,\quad k=1,2\,.
\label{Omega}
\end{equation}
To eliminate zero modes at least two lattice directions need to be ``twisted''. In practice, one can either
explicitely implement the twist eq.~(\ref{Utwist}) or multiply plaquettes in the corners of twisted planes with suitable phase
factors~$\eta,\eta^*$, otherwise maintaining PBC. We opted for the first method. The effect of TBC is twofold:
first, TBC automatically eliminate the undesired zero modes. Second, they drastically reduce finite lattice
size effects: for a given number of~$L^4$ lattice points,
TBC restrict the possible gluon momenta~$p_\nu$ to
 \begin{equation}
   p_\nu = \left\{ \begin{array}{cl}
   \frac{2\pi}{3L} n_\nu , & \nu = \mbox{twisted direction}\,, \\\\
   \frac{2\pi}{L} n_\nu, & \nu = \mbox{periodic direction}\,.
   \end{array} \right.
\label{kQuantize}
\end{equation}
To put it differently, the momenta~$p_\nu$ are quantized as if they lived on a three times bigger lattice for each twisted direction.
\FIGURE{
        \includegraphics[clip,width=.75\textwidth]{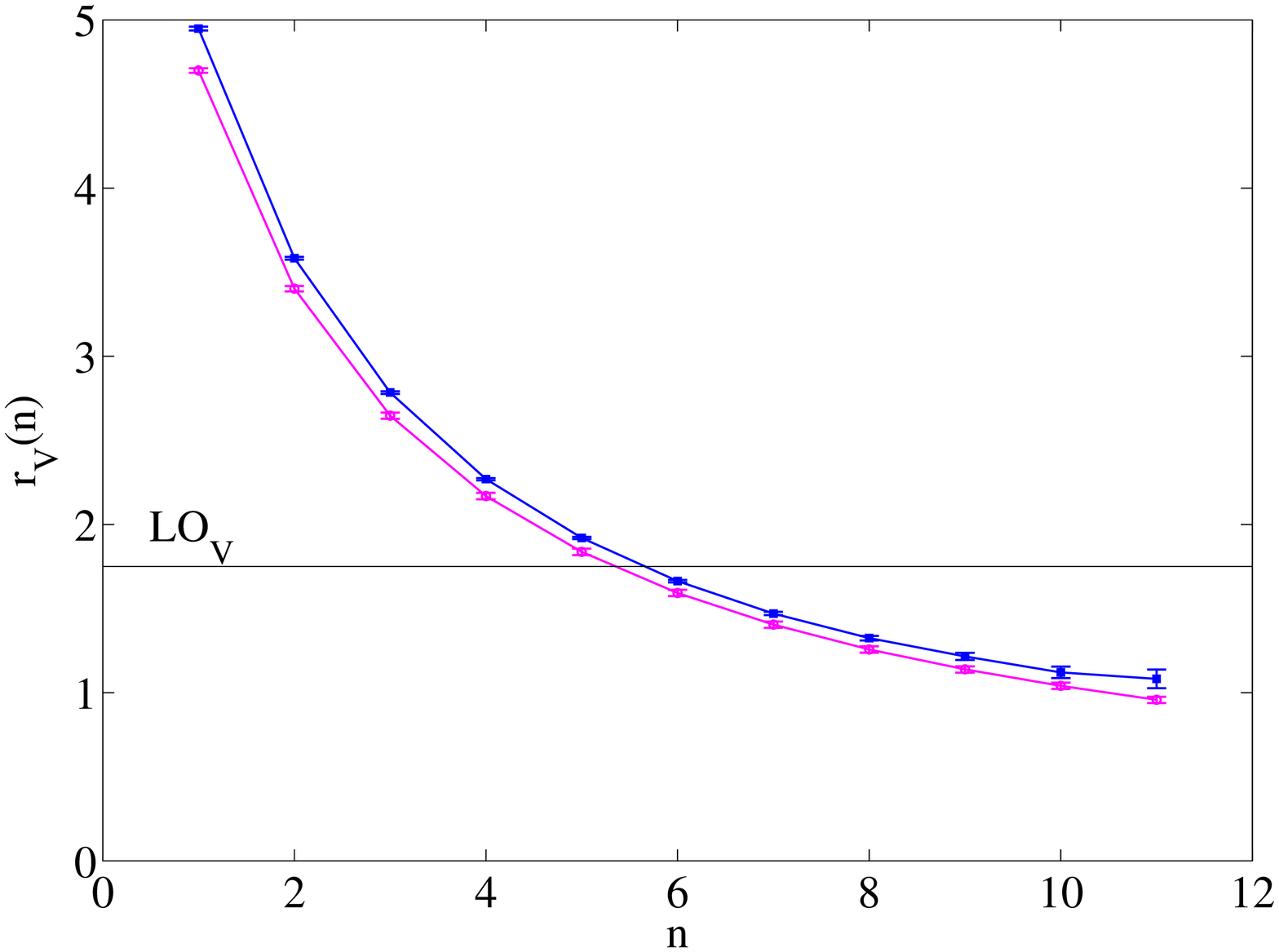}
  \caption{Ratios~$r_{\textrm{v}}(n)$ for the static energy on a $8^4$~lattice, using TBC in three directions (blue squares) and PBC (magenta circles), respectively.
\label{fig:3}}}
\section{Renormalon observables}
So far the only observables that have been checked for a renormalon within
NSPT are the plaquette $\langle U_{\square}\rangle$ and small Wilson loops~\cite{8Loop,VolFin,Rakow:2005yn,Ilgenfritz,Horsley}.
The factorial growth of the coefficients~$w_n$ in the expansion,
\begin{equation}
\label{eq:Plaq_pert}
\langle U_{\square}\rangle=\sum_{n=0}^{\infty} w_n \alpha^{n+1\,},
\end{equation}
translates into the leading-order expectation (see e.g.\
ref.~\cite{Beneke:98} and eq.~(\ref{eq:renormal})),
\begin{equation}
\mathrm{LO_{\mathrm{W}}}=\lim_{n\to\infty}
r_{\textrm{w}}(n):=\lim_{n\to\infty} \frac{|w_{n}|}{n|w_{n-1}|}= 
|a_4|=\frac{11}{8\pi}\,.
%\mathrm{LO_{\mathrm{W}}}=\lim_{n\to\infty} \frac{w_{n}}{w_{n-1}}\, \frac{1}{n}= \frac{11}{8\pi}.
%\textrm{NLO: } \lim_{n\to\infty} c_{n}/c_{n-1} &= \frac{11}{8\pi}n\left(1+\frac{102}{242n}\right)
\end{equation}
The static energy~$V_{\mathrm{self}}$ which we focus on
can be extracted from gauge-invariant Polyakov loop expectation values
$\langle P\rangle$ wrapping around the $T$~direction of an
$L^3\cdot T$ lattice:
\begin{eqnarray}
 V_{\mathrm{self}}=\lim_{T\to\infty}\left(-\frac{1}{T}\mathrm{ln}\, \langle P \rangle\right)\,.
%V_{\mathrm{self}}&=\sum  V^{(n)}_{\mathrm{self}}\alpha^n
 %V_{\mathrm self}\equiv \delta m.
\label{eq:vself}
\end{eqnarray}
The self-energy of a static quark is linearly UV-divergent. Hence,
the perturbative coefficients~$v_n$ of the expansion of $V_{\mathrm{self}}$
are expected to be sensitive to a leading UV renormalon at $u=1/2$:
the leading-order expectation reads~\cite{Beneke:94,Beneke:95,Pineda:01},
 \begin{equation}
\mathrm{LO_{\mathrm{V}}}=\lim_{n\to\infty}
r_{\textrm{v}}(n)=\lim_{n\to\infty} \frac{|v_{n}|}{n|v_{n-1}|}= \frac{11}{2\pi}= 4\,\mathrm{LO_{\mathrm{W}}},
%\mathrm{LO_{\mathrm{V}}}=\lim_{n\to\infty} \frac{v_{n}}{v_{n-1}}\, \frac{1}{n}= \frac{11}{2\pi}= 4\,\mathrm{LO_{\mathrm{W}}},
%\textrm{NLO: } \lim_{n\to\infty} c_{n}/c_{n-1} &= \frac{11}{8\pi}n\left(1+\frac{102}{242n}\right)\,.
\label{eq:LOV}
\end{equation}
\section{Preliminary results} 
Ref.~\cite{Nobes} triggered our interest in combining the static energy
calculation with TBC. In this reference the static energy was calculated
for various lattice sizes at first and second order. TBC in three spatial
directions (TBC3) and even more so TBC in two spatial directions (TBC2) were
found to approach the infinite-volume values much faster than PBC.
We ran simulations up to~$O(\alpha^{12})$ and confirm these findings
at higher orders. In fig.~\ref{fig:3} we employ both TBC3 and PBC to
calculate the static energy on an $8^4$~lattice volume, resulting
in two sets of ratios~$r_{\textrm{v}}(n)$, see eq.~(\ref{eq:LOV}).
For large~$n$, the TBC3 ratios lie significantly closer
to~$\mathrm{LO_{\mathrm{V}}}$ than the PBC ratios.
 
\FIGURE{
        \includegraphics[clip,width=.75\textwidth]
         {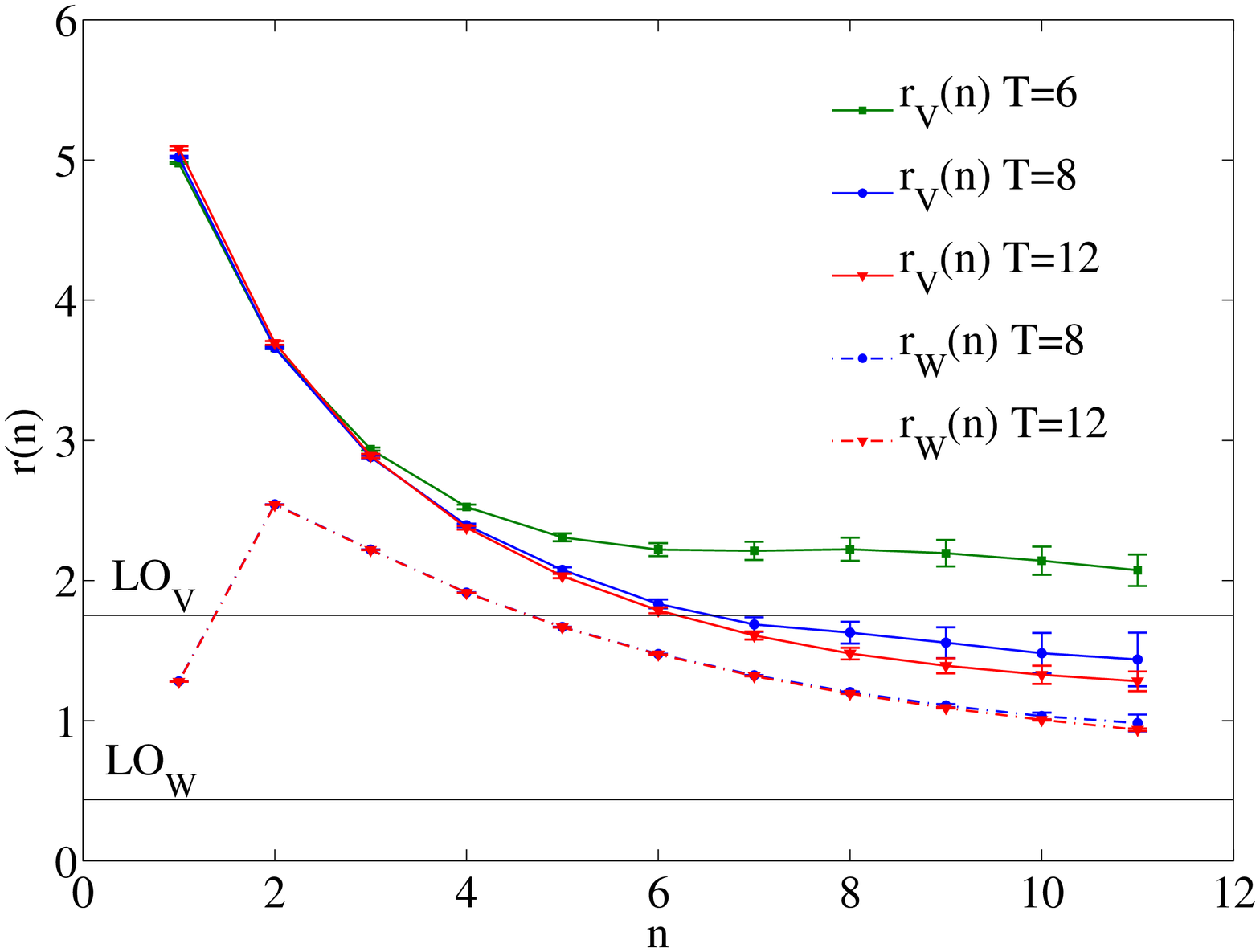}
  \caption{Ratios~$r_{\textrm{v}}(n)$ for the static energy from lattice volumes~$6^4$~(green sqares), $6^3\cdot8$~(blue circles) and $6^3\cdot12$~(red triangles). For the latter two lattice volumes also the plaquette ratios $r_{\textrm{w}}(n)$ are shown (dashed-dotted lines).
  \label{fig:1}}}
We kept the spatial volume fixed to $L^3=6^3$ to test the
viability of eq.~\eqref{eq:vself} at finite~$T=6,8,12$.
Fig.~\ref{fig:1} illustrates that the ratio curve drops
significantly when increasing $T$ from $T=6$~to $T=8$.
Obviously, $T=6$ does not yet probe the large-$T$ limit.
In contrast, the~$T=12$ ratios agree within errors with the $T=8$~data,
indicating the onset of convergence towards the static energy and
its renormalon. Fig.~\ref{fig:1} also includes the plaquette ratios
for~$T=8$ and $T=12$ and these practically coincide. This milder
volume dependence for this more localized quantity seems very plausible.
We point out the clear separation between plaquette and static energy
ratios. Since the renormalon dominance of $V_{\mathrm{self}}$
only starts around the order $n\approx 8$, we would not expect the
plaquette ratios to saturate at their asymptotic value for $n<30$.

We also implemented stout smearing for the temporal links
(once, with smearing parameter $\rho=1/6$) and calculated
eq.~\eqref{eq:vself} in the adjoint representation.
The outcome is presented in fig.~\ref{fig:2} for the TBC2
simulation on the $6^3\cdot12$ volume. We find that, as far
as a potential renormalon is concerned, smearing only affects
low~($n=1,2$) perturbative orders, while higher-order ratios
collapse onto the unsmeared values. Similarly, the change
in representation makes no difference regarding the renormalon
position. The adjoint coefficients at large orders are
also interesting in view of Casimir scaling
violations~\cite{Anzai:2010td,Bali:2003jq}.
\FIGURE{
\includegraphics[clip,width=.75\textwidth]{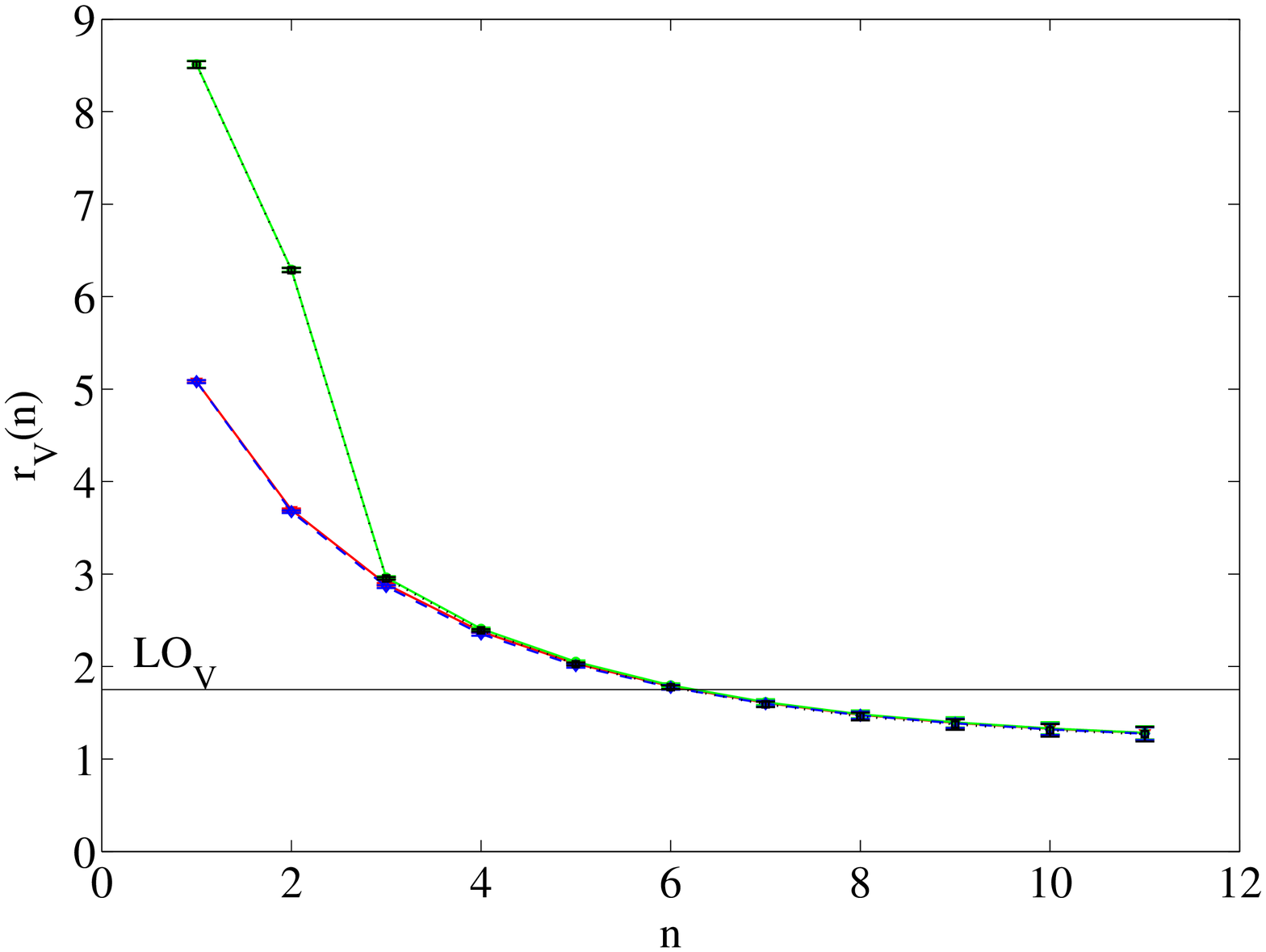}
  \caption{Ratios~$r_{\textrm{v}}(n)$ on a $6^3\cdot12$~lattice. Again we plot the static energy, with (green, solid) and without stout smearing (red,solid). In addition, the static energy in the adjoint representation is shown with (black, dotted) and without smearing (blue, dashed).
  \label{fig:2}}}
\section{Summary}
The perturbative static energy is expected to sense a leading
renormalon emerging four times faster than its plaquette counterpart.
In an exploratory study we have calculated the static energy from
Polyakov loops in NSPT up to~$O(\alpha^{12})$ on small lattice
volumes, where the use of TBC has proven to drastically reduce
finite-size effects. Given the lattice sizes we used and the fact
that theoretical predictions are within range, we are confident
that our ongoing large-volume simulations will shed more light on the
static energy renormalon. 
\acknowledgments
We thank Antonio Pineda for teaching us renormalon theory and
Margarita Garc\'{\i}a P\'erez, Christian Torrero, Howard Trottier and Arwed Schiller for discussions.
We thank the LRZ Munich for computer time. Computations
were also performed on Regensburg's Athene HPC cluster.
C.B.\
is grateful for support from the Studienstiftung des deutschen
Volkes and from the Daimler und Benz Stiftung.
This work was supported by DAAD (Acciones Integradas
Hispano-Alemanas D/07/13355),
DFG (Sonderforschungsbereich/Transregio 55) and the
EU (grants 238353, ITN STRONGnet
and 227431, HadronPhysics2).

\end{document}